\documentclass[aps,prb,reprint,superscriptaddress,noshowkeys]{revtex4-2}
\usepackage{graphicx,dcolumn,bm,microtype,multirow,amscd,amsmath,amssymb,amsfonts,mathtools,physics,wrapfig,siunitx,mleftright,bbold,xspace}

\usepackage[version=4]{mhchem}
\usepackage[normalem]{ulem}
\usepackage[dvipsnames]{xcolor}
\usepackage[utf8]{inputenc}
\usepackage[T1]{fontenc}

\usepackage{hyperref}
\hypersetup{
    colorlinks,
    linkcolor={red!50!black},
    citecolor={red!70!black},
    urlcolor={red!80!black}
}
\usepackage{cleveref}

\usepackage{tikz}
\usepackage{tikz-feynman}
\usepackage{tikz-feynhand}

\newcommand{\titou}[1]{\textcolor{black}{#1}}

\newcommand{\pushright}[1]{\ifmeasuring@#1\else\omit\hfill$\displaystyle#1$\fi\ignorespaces}
\newcommand{\SupInf}{\textcolor{blue}{Supplemental Material}\xspace}

\newcommand{\LCPQ}{Laboratoire de Chimie et Physique Quantiques (UMR 5626), Universit\'e de Toulouse, CNRS, UPS, France}
\newcommand{\LPT}{Laboratoire de Physique Th\'eorique, Universit\'e de Toulouse, CNRS, UPS, France}
\newcommand{\ETSF}{European Theoretical Spectroscopy Facility (ETSF)}

\begin{document}	

\title{Anomalous propagators and the particle-particle channel: Hedin's equations}

\author{Antoine \surname{Marie}}
       \email{amarie@irsamc.ups-tlse.fr}
       \affiliation{\LCPQ}

\author{Pina \surname{Romaniello}}
       \affiliation{\LPT}
       \affiliation{\ETSF}
       
\author{Pierre-Fran\c{c}ois \surname{Loos}}
       \email{loos@irsamc.ups-tlse.fr}
       \affiliation{\LCPQ}

\begin{abstract}
  Hedin's equations provide an elegant route to compute the exact one-body Green's function (or propagator) via the self-consistent iteration of a set of non-linear equations.
  Its first-order approximation, known as $GW$, corresponds to a resummation of ring diagrams and has shown to be extremely successful in physics and chemistry. Systematic improvement is possible, although challenging, via the introduction of vertex corrections.
  Considering anomalous propagators and an external pairing potential, we derive a new self-consistent set of closed equations equivalent to the famous Hedin equations but having as a first-order approximation the particle-particle (pp) $T$-matrix approximation where one performs a resummation of the ladder diagrams.
  This pp version of Hedin's equations offers a way to go systematically beyond the $T$-matrix approximation by accounting for low-order pp vertex corrections.
\end{abstract}

\maketitle

\section{Resummation in many-body perturbation theory}
\label{sec:intro}

In 1965, Lars Hedin published a seminal paper that introduced a set of equations,
\begin{subequations}
  \label{eq:hedins_eq}
  \begin{align}
    G(11') & = G_0(11') + G_0(12)\Sigma(22')G(2'1'), 
    \\
    \Sigma_{\text{xc}}(11') & = \ii G(33') W(12';32) \tilde{\Gamma}(3'2;1'2'), 
    \\
    \begin{split}
    W(12;1'2') 
    & = v(1 2^-;1' 2') 
    \\
    & - \ii W(14;1'4') \tilde{L}(3'4';3^+4) v(2 3;2' 3'),  
    \end{split}
    \\
    \tilde{L}(12;1'2') & = G(13)G(3'1')\tilde{\Gamma}(32;3'2'), 
    \\
    \begin{split}
    \tilde{\Gamma}(12;1'2') & = \delta(12')\delta(1'2) 
    \\
    & + \Xi_{\text{xc}}(13';1'3) G(34)G(4'3') \tilde{\Gamma}(42;4'2'),
    \end{split}
  \end{align}
\end{subequations}
now referred to as Hedin's set \cite{Hedin_1965}.
The composite index 1 gathers time, spin, and spatial variables, and implicit integration over repeated indices is assumed.
The quantities involved will be defined in the following discussion.
\titou{Here, we considered the 4-point version of these equations \cite{Starke_2012,Maggio_2017b,Orlando_2023b}. 
(The form of the present set is not exactly the one introduced by Hedin in Ref.~\onlinecite{Hedin_1965} but it is more convenient in the context of this work.)}
This set of non-linear equations provides a theoretical route to compute the exact one-body Green's function (or propagator) $G$ through self-consistent iterations.

While this formal recipe to obtain the exact propagator is elegant, the main success of Hedin's equations lies in its first-order approximation, the so-called $GW$ approximation \cite{Martin_2016,Reining_2018,Golze_2019,Marie_2023a}, which is achieved by discarding the second term of the 4-point irreducible vertex function $\tilde{\Gamma}$.
Hence, the $GW$ exchange-correlation (xc) self-energy,
\begin{equation}
  \label{eq:gw_selfenergy}
  \Sigma_{\text{xc}}^{GW}(11') = \ii G(22')W(2'1;1'2),
\end{equation}
is expressed in terms of the one-body propagator and the dynamically-screened Coulomb interaction $W$.
The $GW$ approximation has been first employed to compute the photoemission spectrum of solids \cite{Strinati_1980,Strinati_1982a,Strinati_1982b,Strinati_1988,Hybertsen_1985,Hybertsen_1986,Godby_1986,Godby_1987,Godby_1987a,Godby_1988,Blase_1995} before being imported in quantum chemistry to calculate electron attachment and detachment energies in molecular systems \cite{Grossman_2001,Rostgaard_2010,Ke_2011,Blase_2011b,Korzdorfer_2012,Sharifzadeh_2012a,Bruneval_2012,Bruneval_2013,vanSetten_2013,Koval_2014,Knight_2016,Bruneval_2021}. It has proven to be successful in both fields for weakly and moderately correlated systems.

\begin{figure}
  \centering
  \includegraphics[width=\linewidth]{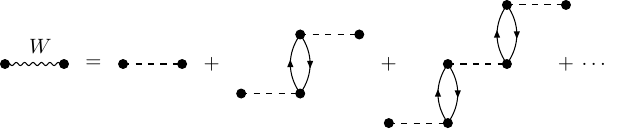}
  \caption{The dynamically-screened interaction $W$ (wiggly line) computed at the ph-RPA level corresponds to a resummation of bubble diagrams. The dashed lines represent the Coulomb interaction and the solid lines with arrows denote the one-body propagator.}
   \label{fig:screened_w}
\end{figure}

The $GW$ self-energy defined in Eq.~\eqref{eq:gw_selfenergy} corresponds to the first-order term of an expansion with respect to the effective interaction $W$.
While the associated dynamical screening can theoretically be computed using any irreducible particle-hole (ph) correlation function $\tilde{L}$, Hedin's equations naturally suggest relying on the same approximated vertex function in $\tilde{L}$ and $\Sigma$.
If this choice is made, the ph direct random-phase approximation (RPA) polarizability \cite{Schuck_Book} naturally appears in the construction of the screened interaction.
The diagrammatic content of the corresponding effective interaction is illustrated in Fig.~\ref{fig:screened_w}.
The ph-RPA is well-known as it corresponds to the resummation of polarizability diagrams that are the most important in the uniform electron gas at high density, the so-called bubble (or ring) diagrams \cite{Bohm_1951,Pines_1952,Bohm_1953,Nozieres_1958,Gell-Mann_1957,MattuckBook}.

On the other hand, the most relevant diagrams in the low-density limit of the uniform electron gas with short-range interactions (as well as in nuclear matter) are quite different \cite{MattuckBook}.
These diagrams, known as ladders, and their exchange counterparts can also be resummed and this yields the analog particle-particle (pp) RPA, also known as pairing vibration approximation \cite{Schuck_Book}.
These two closely related approximations include two different types of correlation events and, hence, do not yield the same correlation energies.
The pp-RPA correlation energy has been shown to be equivalent to coupled cluster with double excitations (CCD) restricted to ladder terms \cite{Scuseria_2013,Peng_2013}.
Similarly, the ph-RPA correlation energy is equivalent to CCD restricted to another subset of terms, namely the ring terms \cite{Scuseria_2008}.

As mentioned earlier, the ph-RPA appears naturally in the $GW$ approximation.
Within Hedin's equations, ladder self-energy diagrams are obtained through vertex corrections. 
At each self-consistent iteration of Eqs.~\eqref{eq:hedins_eq}, the functional derivative of $W$ appearing in $\tilde{\Gamma}$ (through the exchange-correlation kernel $\Xi_\xc =  \fdv*{\Sigma_\xc}{G}$) creates an additional ladder self-energy diagram of higher order \cite{Schindlmayr_1998,Mejuto-Zaera_2022}.
However, in certain physical situations, it is preferable to account for all ladder diagrams from the start.
This is achieved by the $T$-matrix approximation.

\begin{figure}
  \centering
  \includegraphics[width=\linewidth]{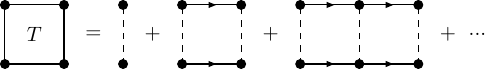}
  \caption{The effective interaction $T$ computed at the pp-RPA level corresponds to a resummation of ladder diagrams. The dashed lines represent the Coulomb interaction and the solid lines with arrows denote the one-body propagator. The exchange counterpart of each of these diagrams should be also included but has not been represented here.}
   \label{fig:effective_t}
\end{figure}

The $T$-matrix approximation \cite{Baym_1961,Danielewicz_1984a}, also known as the Bethe-Goldstone approximation \cite{Bethe_1957}, has first been introduced in the nuclear many-body problem \cite{FetterBook,Dickhoff_2008}.
The $T$-matrix is a 4-point effective interaction accounting for repeated scattering of two particles. In practice, these scattering events are often computed at the pp-RPA level \cite{Schuck_Book}.
This is in close analogy with the effective interaction $W$ accounting for screening events and built using the ph-RPA.
The diagrams resummed in the $T$-matrix effective interaction are represented in Fig.~\ref{fig:effective_t}.
Note that the term $T$-matrix has been employed in various contexts for different types of effective interaction and they should not be confused (see, for example, the electron-hole $T$-matrix for electron-magnon scattering \cite{Romaniello_2012,Muller_2019,Nabok_2021,Orlando_2023b}).
The $T$-matrix-based self-energy has been applied to model systems, like the Hubbard model \cite{Romaniello_2012,Gukelberger_2015}, as well as more realistic solids (though often combined with other correlation channels in this case) \cite{Liebsch_1981,Springer_1998,Katsnelson_1999,Katsnelson_2002,Zhukov_2005}.
One of its main successes in this field is the description of the \SI{6}{\eV} satellite of nickel \cite{Liebsch_1981,Springer_1998}.
More recently, it has been used to compute ionization potentials of molecular systems \cite{Zhang_2017,Li_2021b,Li_2023,Monino_2023,Orlando_2023b,Marie_2024}, where it has been shown to have similar accuracy to $GW$ for valence ionization potentials if a Hartree-Fock reference is employed for both \cite{Zhang_2017}.
\titou{Finally, the $T$-matrix has also been applied in various other fields, especially those where pairing correlations are important, such as nuclear matter \cite{Bozek_1999,Soma_2006,Soma_2008,Dickhoff_2008}, superconducting materials \cite{Pieri_2004,Chen_2005,Sopik_2011,Tajima_2020,Zeng_2022}, or Fermi gases \cite{Perali_2002,Chen_2005,Pini_2019}.}

Unfortunately, while Hedin's equations provide a path to go beyond $GW$, to the best of our knowledge, there is no such set of equations for the $T$-matrix approximation.
The $T$-matrix was initially introduced as a resummation of diagrams, or equivalently as a Bethe-Salpeter equation for an effective 4-point interaction \cite{FetterBook}.
On the other hand, Hedin's equations stem from a functional derivative framework \cite{Martin_2016}.
Romaniello \etal~managed to obtain the $T$-matrix in such a framework \cite{Romaniello_2012}.
Their derivation highlights connections between the $GW$ and $T$-matrix approximations, as well as ways to combine them to go beyond $GW$.
However, it does not provide a straightforward pathway for a systematic inclusion of vertex corrections in the $T$-matrix approximation, as in the case of $GW$.
Vertex corrections to the $GW$ self-energy is an active field of research \cite{DelSol_1994,Shirley_1996,Schindlmayr_1998,vanLeeuwen_2006,Shishkin_2007b,Romaniello_2009a,Romaniello_2012,Gruneis_2014,Maggio_2017b,Lewis_2019,Mejuto-Zaera_2021,Wang_2021a,Forster_2022a,Mejuto-Zaera_2022,Weng_2023,Bruneval_2024}, and extending these corrections to the $T$-matrix approximation would undoubtedly offer valuable new insights.

The primary focus of the present manuscript is to bridge this gap by deriving, from first principles, an alternative set of equations for the one-body propagator.
Within this novel framework, the $T$-matrix emerges naturally through the lowest-order vertex approximation, in close analogy with the $GW$ approximation. 
Therefore, we shall refer to it as the pp version of Hedin's equations.
The crux of the derivation lies in the consideration of anomalous propagators and a non-number-conserving external potential, as elaborated in the subsequent sections.
The present work aligns with \titou{several} studies dealing with the generalization of Hedin's equations to a spin-dependent interaction \cite{Aryasetiawan_2008}, the exploration of connections between the parquet and $GW\Gamma$ formalisms \cite{Krien_2021}, or the extension of Hedin's equations to the Gorkov propagator \cite{PhDEssenberger}.

\section{Self-Energy and Schwinger relations}
\label{sec:self_energy}

The central object of this closed set of equations is the equilibrium time-ordered one-body propagator (at zero temperature) defined as
\begin{equation}
  \label{eq:one_body_propag}
  G(11') = (-\ii)\mel{\Psi_0^N}{\Hat{T}\mqty[\hpsi(1)\hpsid(1')]}{\Psi_0^N},
\end{equation}
where $\ket{\Psi_0^N}$ is the exact $N$-electron ground-state wave function. The time-ordering operator $\Hat{T}$ acts on the annihilation and creation field operators in the Heisenberg picture, which read
\begin{subequations}
\begin{align}
  \label{eq:heisenberg_a}
  \hpsi(1)  &= \hpsi(\bx_1,t_1)   = e^{\ii\Hat{H}t_1} \Hat{\psi}(\bx_1)  e^{-\ii\Hat{H}t_1}, 
  \\
  \label{eq:heisenberg_b}
  \hpsid(1) &= \hpsid(\bx_1,t_1)  = e^{\ii\Hat{H}t_1} \Hat{\psi}^\dagger(\bx_1) e^{-\ii\Hat{H}t_1},
\end{align}
\end{subequations}
where $\hH$ is the electronic Hamiltonian and $\bx_1$ is a variable gathering spin and position $\br_1$.

The first step is the same as in the usual derivation of Hedin's equations (see, for example, Ref.~\onlinecite{Strinati_1988}) and consists of deriving the Dyson equation,
\begin{equation}
  \label{eq:dyson_eq}
  G(11') = G_0(11') + G_0(12)\Sigma(22')G(2'1'),
\end{equation}
from the equation of motion for $G$.
Here, $G_0$ is the non-interacting one-body propagator and the self-energy is defined as
\begin{equation}
  \label{eq:self_energy}
  \Sigma(11') =  -\ii v(12;3'2') G_2(2'^+3';2^{++}3) G^{-1}(31').
\end{equation}
This definition involves the inverse of the one-body propagator $G^{-1}$, the 4-point Coulomb interaction
\begin{equation}
  \label{eq:4point_coulomb}
  v(12;1'2') = \delta(11')v(12)\delta(22')
\end{equation}
with
\begin{equation}
  \label{eq:2point_coulomb}
  v(12) = \frac{\delta(t_1 - t_2)}{\abs{\br_1-\br_2}},
\end{equation}
and the two-body propagator
\begin{equation}
  \label{eq:two_body_propag}
  G_2(12;1'2') = (-\ii)^2\mel{\Psi_0^N}{\Hat{T}[\hpsi(1)\hpsi(2)\hpsid(2')\hpsid(1')]}{\Psi_0^N}.
\end{equation}
The notation $1^{\pm}$ means that an infinitesimal shift is added/subtracted to the corresponding time variable and $\delta(11')$ is the Dirac delta function.

To obtain a closed set of equations for $G$, the two-body Green's function must be expressed in terms of the one-body propagator.
This is achieved thanks to the Schwinger relation \cite{Martin_1959}
\begin{equation}
  \label{eq:eh_schwinger}
  G_2(12;1'2') = - \fdv{G(11')}{U^\eh(2'2)} + G(11')G(22'),
\end{equation}
which express $G_2$ in terms of $G$ and its derivative with respect to an electron-hole (eh) external potential $U^\eh$ which is linked to the external operator 
\begin{equation}
  \label{eq:eh_ext_potential}
  \Hat{\mathcal{U}}^\eh(t_2) = \int \dd(\bx_2\bx_{2'}) \hpsid(\bx_2) U^\eh(\bx_2\bx_{2'};t_2) \hpsi(\bx_{2'}).
\end{equation}
as $U^\eh(11') = U^\eh(\bx_{1}\bx_{1'};t_1)\delta(t_{1} - t_{1'})$.
Note that, in Eq.~\eqref{eq:eh_schwinger}, the one- and two-body propagators have been generalized to their non-equilibrium version.
In the following, the functional $U$-dependence of these propagators is not explicitly written for the sake of conciseness.
In the presence of such an external potential, the field operators of Eqs.~\eqref{eq:heisenberg_a} and \eqref{eq:heisenberg_b} have to be generalized to the case of a time-dependent Hamiltonian.
Hence, as explained in detail in the \SupInf, the derivation of the Schwinger relation is more conveniently performed in the interaction representation.

The key idea to obtain an alternative system of equations is to realize that an analog relationship can be obtained in the case of an external pairing potential operator,
\begin{multline}
  \label{eq:eh_ext_potential}
  \Hat{\mathcal{U}}^\pp(t_{2}) = \frac{1}{2} \int \dd(\bx_2\bx_{2'}) 
  \Big[ \hpsid(\bx_{2}) U^\ee(\bx_2\bx_{2'};t_2) \hpsid(\bx_{2'}) 
  \\ + \hpsi(\bx_{2}) U^\hh(\bx_2\bx_{2'};t_2) \hpsi(\bx_{2'}) \Big],
\end{multline}
composed by an electron-electron (ee) and a hole-hole (hh) external potential, $U^\ee$ and $U^\hh$, respectively.
One major difference with $\hat{\mathcal{U}}^\eh$ is that $\Hat{\mathcal{U}}^\pp$ does not commute with the particle number operator.
Therefore, the number of particles is not a good quantum number of the Hamiltonian in the presence of $\hat{\mathcal{U}}^\pp$.
Equivalently, one may say that $\Hat{\mathcal{U}}^\pp$ breaks the $U(1)$ symmetry of the Hamiltonian \cite{MattuckBook}.

The linear response of $G$ to this external perturbation is not linked to $G_2$ as in Eq.~\eqref{eq:eh_schwinger}.
However, $G_2$ can be obtained as the response of an anomalous propagator to this external pairing potential
\begin{equation}
  \label{eq:pp_schwinger}
  G_2(12;1'2') = -2 \fdv{G^\ee(1'2')}{U^\hh(12)} - G^\ee(1'2') G^\hh(12).
\end{equation}
The derivation of this equation closely follows the one of Eq.~\eqref{eq:eh_schwinger} and is reported in the \SupInf.

The anomalous propagator $G^\ee$, and its counterpart $G^\hh$, also known as pairing propagators, are defined as 
\begin{subequations}
  \begin{align}
    \label{eq:anom_propag_ee}
    G^\ee(11') &= (-\ii)\mel{\Psi_0}{\Hat{T}\mqty[\hpsid(1)\hpsid(1')]}{\Psi_0}, 
    \\
    \label{eq:anom_propag_hh}
    G^\hh(11') &= (-\ii)\mel{\Psi_0}{ \Hat{T}\mqty[\hpsi(1)\hpsi(1')] }{\Psi_0}.
  \end{align}
\end{subequations}
Therefore, the one-body propagator defined in Eq.~\eqref{eq:one_body_propag} will now be denoted as $G^\he$ and referred to as the normal propagator. 
Thanks to Nambu's matrix formalism \cite{Nambu_1960}, these propagators can be gathered in a single entity
\begin{equation}
  \label{eq:gorkov_propag}
  \bG(11') = \mqty(G^\he(11') & G^\hh(11') \\ G^\ee(11') & G^\eh(11')),
\end{equation}
known as the Gorkov propagator \cite{Gorkov_1958}.
The lower-right eh propagator is linked to the normal propagator by the relationship $G^\eh(11') = -G^\he(1'1)$.
The Gorkov propagator admits a Dyson equation
\begin{equation}
  \label{eq:gorkov_dyson}
  \bG(11') = \bG_0(11') + \bG_0(12) [\bSig(22')+\bU(22')] \bG(2'1'),
\end{equation}
which defines the normal and anomalous components of the corresponding self-energy in Nambu's formalism
\begin{equation}
  \begin{split}
  \label{eq:se_nambu}
  \bSig(11') &= \mqty(\Sigma^\he(11') & \Sigma^\hh(11') \\ \Sigma^\ee(11') & \Sigma^\eh(11')) \\
             &= \bG_0^{-1}(11') - \bG^{-1}(11') - \bU(11').
  \end{split}
\end{equation}
The matrix $\bG_0$ is the independent-particle Gorkov propagator and
\begin{equation}
  \bU(11') = \mqty( 0 & U^\ee(11') \\ U^\hh(11') & 0 ).
\end{equation}
Note that $U^\ee$ appears in the hh component of Nambu's matrix formalism and vice-versa.
This property is a direct consequence of the equation of motion for $\bG$ which is derived in the \SupInf.

A diagrammatic perturbation expansion of $\bSig$ in terms of the Coulomb interaction exists as in the simpler case of $\Sigma^\he$ \cite{Nambu_1960,MattuckBook}.
Recently, this perturbation expansion has been derived up to second order, implemented, and applied to mid-mass nuclei in the context of nuclear structure calculations \cite{Soma_2011,Soma_2013,Soma_2014,Soma_2020,Soma_2021,Porro_2021}. (See also Ref.~\onlinecite{Barbieri_2022} for a recent extension of the Gorkov algebraic diagrammatic construction up to third order.)

Note that, in the definition of $G^\ee$ and $G^\hh$, the superscript $N$ characterizing the ground-state wave function has been removed.
Indeed, as mentioned above, the number of particles is not conserved in the presence of the external pairing potential. Hence, the wave function becomes a linear superposition of wave functions with various particle numbers. 
If a wave function with a fixed number of particles is considered, then the anomalous propagators vanish [see Eqs.~\eqref{eq:anom_propag_ee} and \eqref{eq:anom_propag_hh}].
\titou{For the non-relativistic electronic Hamiltonian, this will always be the case for the exact wave function of a finite system $\ket{\Psi_0^N}$ as this Hamiltonian does not spontaneously break the particle-number symmetry \cite{Bach_1994,BlaizotBook,vanAggelen_2014}.
In some cases, such as superconductivity or nuclear superfluidity \cite{Bardeen_1957,Dean_2003}, relying on symmetry-broken approximate wave functions and the associated non-zero anomalous propagators is essential to describe the physics at play.}

At first, it might seem counterintuitive to use $G^\ee$ and $G^\hh$ with a number-conserving Hamiltonian.
However, it is crucial to realize that while anomalous quantities are zero when the pairing potential is switched off, their derivatives with respect to the pairing external potential can be non-zero at $U=0$.
This is exemplified by taking the equilibrium limit of Eq.~\eqref{eq:pp_schwinger} where the derivative of $G^\ee$ with respect to $U^\hh$ evaluated at $U=0$ is equal to $G_2$.

Before going further, we should mention that anomalous quantities and/or pairing potentials have also been explored in various ways in quantum chemistry \cite{Staroverov_2002,Tsuchimochi_2009,Scuseria_2009,Tsuchimochi_2010a,Tsuchimochi_2010,Ellis_2011,Nishida_2023,Matveeva_2023,Matveeva_2024}.
One directly related example is the work of Yang's group on pairing fields in density-functional theory (DFT) \cite{vanAggelen_2013,vanAggelen_2014,Peng_2014b}.
They formulated the adiabatic connection fluctuation dissipation theorem in terms of pairing matrix fluctuations which leads to a new path to develop density functional approximations \cite{vanAggelen_2013,vanAggelen_2014}.
They also extended the adiabatic time-dependent DFT (TDDFT) formalism to an external pairing field \cite{Peng_2014b}.
This alternative response problem is closely related to pp-RPA and yields complementary information to the usual ph-TDDFT problem.
Another example is the variation-after-projection ansatz where the particle-number symmetry of a Hartree-Fock determinant is restored before variational optimization at a mean-field cost \cite{Scuseria_2011} (see also Ref.~\onlinecite{Lacroix_2012}).
Finally, Johnson and co-workers employed Richardson-Gaudin states (the eigenfunctions of the Bardeen-Cooper-Schriffer model Hamiltonian \cite{Bardeen_1957}) as building blocks to describe strongly correlated molecular systems \cite{Johnson_2020,Fecteau_2021,Fecteau_2022,Johnson_2023a,Johnson_2023}.

\section{Particle-Particle Gorkov-Hedin Equations}
\label{sec:pp_gorkov_hedin}

The stage is now set to derive the pp version of Hedin's equations.
As mentioned earlier, the relevant equations for a number-conserving Hamiltonian are the ones involving only $G^\he$ and $\Sigma^\he$.
However, because the Schwinger relation involves the other components of the Gorkov propagator, it is more convenient to derive a closed set of equations for $\bG$ (at finite $\bU$), hence referred to as the pp Gorkov-Hedin equations.
Then, the equations relevant for number-conserving Hamiltonians are recovered in the limit of a vanishing pairing potential.
This will be done in Sec.~\ref{sec:pp_hedin} where the link with the conventional Hedin equations will be discussed.
In this section, an overview of the derivation of the pp Gorkov-Hedin equations is provided.
A more comprehensive derivation can be found in the accompanying \SupInf.

\begin{widetext}

As mentioned earlier, the Gorkov-Dyson equation can be derived from the equation of motion for $\bG$.
The resulting self-energy expression is
\begin{equation}
  \label{eq:eom_nambu_self_nrj}
  \bSig(11') = - \ii \mqty( v(12^{--};32'^{-}) & 0 \\ 0 & - v(32^{+};12'^{++}) ) \bG_2(2'3;23') \bG^{-1}(3'1'),
\end{equation}
where $\bG_2$ is a Nambu generalization of the two-body Green's function
\begin{equation}
  \bG_2(12;1'2') = (-\ii)^2 \mel{\Psi_0}{\hT\left[\mqty(\hpsi(1)\hpsi(2)\hpsid(2')\hpsid(1') & \hpsi(1)\hpsi(2)\hpsi(2')\hpsid(1') \\ \hpsi(1)\hpsid(2)\hpsid(2')\hpsid(1') & \hpsi(1)\hpsid(2)\hpsi(2')\hpsid(1') ) \right]}{\Psi_0}.
\end{equation}
The Schwinger relation of Eq.~\eqref{eq:pp_schwinger} can be extended to $\bG_2$ in order to obtain a closed set of equations for $\bG$
\begin{equation}
  \bG_2(12;1'2') =
  \begin{pmatrix}
    - 2 \dfrac{\delta G^{\ee}(1'2')}{\delta U^\hh(12)} - G^{\ee}(1'2')G^{\hh}(12) & - 2 \dfrac{\delta G^{\eh}(1'2')}{\delta U^\hh(12)} - G^{\eh}(1'2')G^{\hh}(12) \\
    - 2 \dfrac{\delta G^{\he}(12')}{\delta U^\ee(21')} -  G^{\he}(12')G^{\ee}(21') & - 2 \dfrac{\delta G^{\hh}(12')}{\delta U^\ee(21')} - G^{\hh}(12')G^{\ee}(21') \end{pmatrix}.
\end{equation}
Substituting this relation into Eq.~\eqref{eq:eom_nambu_self_nrj} leads to two self-energy terms.
The term corresponding to the product of propagators reads
\begin{equation}
  \label{eq:bogol}
  \bSig_{\text{B}}(11') = \ii \mqty(0 & v(11') G^{\hh}(1'^{-}1) \\ v(11') G^{\ee}(1'^{+}1) & 0 ), 
\end{equation}
and is identified as the first-order static anomalous self-energy or Bogoliubov (B) self-energy.
Therefore, the self-energy stemming from the remaining term in the Schwinger relation, denoted as $\bSig_{\text{Hxc}}$, accounts for Hartree (H), exchange (x) and correlation effects (c).

Through the link between the derivative of the Gorkov propagator and the derivative of the inverse Gorkov propagator (see \SupInf), $\bSig_{\text{Hxc}}$ can be expressed as
\begin{multline}
  \label{eq:hxc}
  \bSig_{\text{Hxc}}(11') = 2\ii \mqty( v(12;32') & 0 \\ 0 & - v(32';12) ) \\
  \qty[\mqty( G^{\ee}(2^{++}3') & G^{\eh}(2^{++}3') \\ 0 & 0 ) \boldsymbol{\Gamma}^\hh (2'3;3'1')  + \mqty( 0 & 0 \\ G^{\he}(2^{--}3') & G^{\hh}(2^{--}3') ) \boldsymbol{\Gamma}^\ee (32';3'1') ],
\end{multline}
where the vertex functions have been defined as
\begin{align}
  \bGam^\hh (12;1'2')&=-\fdv{\bG^{-1}(1'2')}{U^\hh(1^{+}2)}, & \bGam^\ee (12;1'2')&=-\fdv{\bG^{-1}(1'2')}{U^\ee(12^{-})}.
\end{align}

The self-energy will now be expressed in terms of effective interactions in order to obtain an analog of Hedin's equations.
Mathematically, this is done through the chain rule with respect to the two anomalous total potentials, namely, $V^\ee = \Sigma^\ee_{\text{B}} + U^\hh$ and $V^\hh = \Sigma^\hh_{\text{B}} + U^\ee$, which yields
\begin{multline}
  \label{eq:sigma_hxc_gamma}
  \bSig_{\text{Hxc}}(11') = \ii\left[ \mqty( G^{\ee}(2^{++}3') & G^{\eh}(2^{++}3') \\ 0 & 0 ) 
    \left\{ T^\he(12;44') \tilde{\bGam}^\ee(44';3'1') + T^\hh(12;44') \tilde{\bGam}^\hh(44';3'1') \right\} \right.\\
  \left. + \mqty( 0 & 0 \\ G^{\he}(2^{--}3') & G^{\hh}(2^{--}3') ) \left\{ T^\ee(12;44') \tilde{\bGam}^\ee(44';3'1') + T^\eh(12;44') \tilde{\bGam}^\hh(44';3'1') \right\} \right],
\end{multline}
where the irreducible vertex functions,
\begin{align}
  \tilde{\bGam}^\hh (12;1'2')&=-\fdv{\bG^{-1}(1'2')}{V^\hh(12)}, 
  & 
  \tilde{\bGam}^\ee (12;1'2')&=-\fdv{\bG^{-1}(1'2')}{V^\ee(12)},
\end{align}
and the effective interaction,
\begin{equation}
  \bT(12;1'2') = \mqty( T^\he(12;1'2') & T^\hh(12;1'2') \\ T^\ee(12;1'2') & T^\eh(12;1'2') ) = 2 \mqty( v(12;33') & 0 \\ 0 & - v(33';12) ) 
  \begin{pmatrix}
   \dfrac{\delta V^\ee(1'2')}{\delta U^\hh(3'^{+}3)} & \dfrac{\delta V^\hh(1'2')}{\delta U^\hh(3'^{+}3)} \\ \dfrac{\delta V^\ee(1'2')}{\delta U^\ee(33'^{-})} & \dfrac{\delta V^\hh(1'2')}{\delta U^\ee(33'^{-})}
  \end{pmatrix},
\end{equation}
have been introduced.

This effective interaction admits a Dyson equation
\begin{equation}
  \bT(12;1'2') =  - \bar{\boldsymbol{V}}(12;1'2') - \bT(12;33') \tilde{\bK}(33';44') \boldsymbol{V}(44'^{+};1'2'^{++}),
\end{equation}
where the kernel $\tilde{\bK}$ is equal to
\begin{equation}
  \tilde{\bK}(12;1'2') = \ii \begin{pmatrix} \dfrac{\delta G^\ee(1'2')}{\delta V^\ee(12)} & \dfrac{\delta G^\hh(1'2')}{\delta V^\ee(12)} \\ \dfrac{\delta G^\ee(1'2')}{\delta V^\hh(12)} & \dfrac{\delta G^\hh(1'2')}{\delta V^\hh(12)} \end{pmatrix} = \ii \begin{pmatrix} [\bG(1'3)\tilde{\bGam}^\ee(12;33')\bG(3'2')]^\ee & [\bG(1'3)\tilde{\bGam}^\ee(12;33')\bG(3'2')]^\hh \\ [\bG(1'3)\tilde{\bGam}^\hh(12;33')\bG(3'2')]^\ee & [\bG(1'3)\tilde{\bGam}^\hh(12;33')\bG(3'2')]^\hh \end{pmatrix}.
\end{equation}
The notation $[\bG\tilde{\bGam}^\ee\bG]^\ee$ stands for the ee block of the product matrix $\bG\tilde{\bGam}^\ee\bG$.
The Coulomb potential $\boldsymbol{V}$ is defined as
\begin{equation}
  \boldsymbol{V}(12^{+};1'2'^{++}) = \mqty(v(12^{+};1'2'^{++}) & 0 \\ 0 & v(1'2'^{--};12^{-})),
\end{equation}
and $\bar{\boldsymbol{V}}$ is its antisymmetric counterpart, \ie, $\bar{\boldsymbol{V}}(12;1'2') = \boldsymbol{V}(12;1'2') -\boldsymbol{V}(12;2'1')$.
Therefore, the irreducible vertex functions appear both in the self-energy [see Eq.~\eqref{eq:sigma_hxc_gamma}] (outer vertex) and in the effective interaction (inner vertex).

Using the lowest-order approximations of $\tilde{\bGam}^\hh$ and $\tilde{\bGam}^\ee$,
\begin{align}
  \tilde{\bGam}^\hh_0(12;1'2')&=\frac{1}{2}\mqty( 0 & \delta(1'1)\delta(2'2) - \delta(1'2)\delta(2'1) \\ 0 & 0 ) & \tilde{\bGam}^\ee_0(12;1'2')&=\frac{1}{2}\mqty( 0 & 0 \\ \delta(1'1)\delta(2'2) - \delta(1'2)\delta(2'1) & 0 ),
\end{align}
the corresponding self-energy becomes
\begin{equation}
  \label{eq:gorkov_gt}
  \bSig_\text{Hxc}(11') = \ii \mqty( G^{\eh}(22')T^\he(12;2'1') & G^{\ee}(22')T^\hh(12;2'1') \\ G^{\hh}(22')T^\ee(12;2'1') & G^{\he}(22') T^\eh(12;2'1') ),
\end{equation}
and the kernel of the $\bT$-matrix Dyson equation reads
\begin{equation}
  \label{eq:pp_ltilde_approx}
  \tilde{\bK}(12;1'2') =  \frac{\ii}{2} \qty[\delta(13)\delta(23') - \delta(13')\delta(23)] \mqty( G^\eh(1'3) G^\he(3'2') & G^\hh(1'3) G^\hh(3'2') \\ G^\ee(1'3) G^\ee(3'2') & G^\he(1'3) G^\eh(3'2') ).
\end{equation}
\end{widetext}

\begin{figure}
  \includegraphics[width=\linewidth]{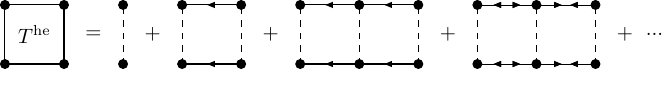}
  \includegraphics[width=\linewidth]{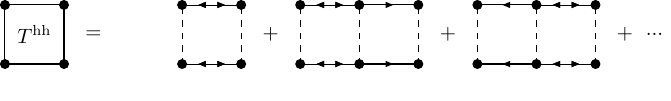}
  \includegraphics[width=\linewidth]{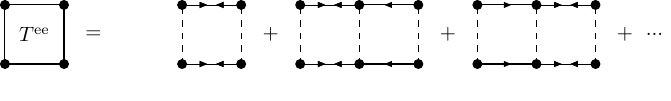}
  \includegraphics[width=\linewidth]{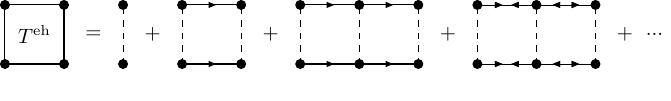}
  \caption{The effective interaction $\bT$ computed with the lowest-order vertex approximation results in a resummation of ladder diagrams for each component. The exchange counterpart of each of these diagrams should also be included but has not been represented here. The double-arrowed propagators in $T^\hh$ ($T^\ee$) represent $G^\hh$ ($G^\ee$) \cite{BlaizotBook}.}
   \label{fig:gorkov_t}
\end{figure}

\titou{The diagrams corresponding to each component of $\bT$ obtained with this kernel are represented, up to third order, in Fig.~\ref{fig:gorkov_t}.
The two anomalous effective interactions have no first-order terms.
This is consistent with the above derivation as the self-energy $\bSig_\text{Hxc}$ does not contain the first-order anomalous self-energies [see Eqs.~\eqref{eq:bogol} and \eqref{eq:hxc}].}
Note that this generalized $T$-matrix approximation was introduced by Bozek using diagrammatic techniques to study superfluid nuclear matter \cite{Bozek_2002}.
Therefore, the present work provides a first-principle derivation of Bozek's $T$-matrix.
It might also be used to go beyond Bozek's approximation by including vertex corrections in $\bT$ and/or $\bSig$.
For example, note that the self-energy approximation of Eqs.~\eqref{eq:gorkov_gt}-\eqref{eq:pp_ltilde_approx} is not complete up to second order in the Coulomb interaction.
The missing second-order terms are recovered through the first iteration of the pp Gorkov-Hedin equations as shown in the \SupInf.
Finally, the extension of conventional Hedin's equations to the Gorkov propagator \cite{PhDEssenberger} has been employed to derive exchange-correlation energy functional for superconducting DFT \cite{Oliveira_1988b,Luders_2005,Marques_2005}.
Hence, the pp Gorkov-Hedin's equations might also provide additional insights into this field.

\section{$T$-Matrix approximation and vertex corrections}
\label{sec:pp_hedin}

Now that the pp Gorkov-Hedin equations have been derived, a set of equations analog to Eqs.~\eqref{eq:hedins_eq} will be recovered as a limiting case ($U \to 0$).
We remind the reader that $G^\he$ is the normal propagator and, thus, we focus on the upper-left block of the Gorkov-Dyson equation.
This leads to the following alternative system of equations
\begin{subequations}
  \begin{align}
    G(11') & = G_0(11') + G_0(12)\Sigma(22')G(2'1'), 
    \\
    \label{eq:sigma_alt_hedin}
    \Sigma(11') & = \ii G(2'2^{++}) T(12;33') \tilde{\Gamma}(33';2'1'),
    \\
    \label{eq:t_alt_hedin}
    \begin{split}
    T(12;1'2') & = - \bar{v}(12;1'2')
    \\
    - &T(12;33') \tilde{K}(33';44') v(44'^{+};1'2'^{++}),
    \end{split}
    \\
    \label{eq:tilde_K_alt_hedin}
    \tilde{K}(12;1'2') & = \ii G(31') G(3'2') \tilde{\Gamma}(12;33'),
    \\
    \label{eq:tilde_gamma_alt_hedin}
    \begin{split}
    \tilde{\Gamma}(12;1'2') & = \frac{1}{2} [ \delta(1'2)\delta(2'1) - \delta(1'1)\delta(2'2) ]
    \\
    - &\Xi^\pp(33';1'2') G(43) G(4'3') \tilde{\Gamma}(12;44'),
    \end{split}
  \end{align}
\end{subequations}
which is actually not closed because the pp kernel,
\begin{equation}
  \Xi^\pp(12;1'2') = \eval{\fdv{\Sigma^\ee(1'2')}{G^\ee(12)}}_{U=0},
\end{equation}
explicitly depends on the anomalous self-energy $\Sigma^\ee$.
Therefore, to compute vertex corrections in this framework, one first needs to compute the corresponding vertex correction for $\bSig$ [see Eq.~\eqref{eq:gorkov_gt}] and then take the number-conserving limit.

\titou{The vertex corrections to $\bSig$ are computed following Mejuto-Zaera and Vl\v{c}ek's procedure for Hedin’s self-consistency \cite{Mejuto-Zaera_2022}. (See also Ref.~\onlinecite{Schindlmayr_1998}.)
Schematically, this is done by starting from a self-energy approximation and computing the associated vertex.
The latter is then inserted back in the self-energy and the effective interaction.
Each iteration, therefore, produces a richer self-energy approximation.
The derivation is explicitly performed in the \SupInf and the corresponding expressions are discussed below.}
\titou{We emphasize that while $G^\ee$ and $G^\hh$ appear in the self-consistency process mentioned above, they vanish in the number-conserving limit.
Therefore, the final expressions of $\Sigma$ and $T$ depend only on $G$ and $v$.
These anomalous propagators are never computed in practice but are only employed as intermediates during the derivation.}

The analog of the $GW$ approximation for this set is obtained by setting the inner and outer vertices to $\tilde{\Gamma}(12;1'2') = \frac{1}{2} [ \delta(1'2)\delta(2'1) - \delta(1'1)\delta(2'2) ]$.
The resulting self-energy is
\begin{equation}
  \label{eq:gt_selfnrj}
  \Sigma(11') = \ii G(2'2^{++})T(12;1'2'),
\end{equation}
with the effective interaction
\begin{multline}
  T(12;1'2') = - \bar{v}(12;1'2') \\
  - T(12;33')K_0(33';44') v(44'^{+};1'2'^{++}),
\end{multline}
and
\begin{equation}
  K_0(12;1'2') = \frac{\ii}{2} \qty[ G(12') G(21') - G(22') G(11') ],
\end{equation}
\titou{where the kernel is recognized to be the non-interacting pp propagator.}
Therefore, this approximate self-energy is exactly the $T$-matrix approximation computed at the pp-RPA level.

While the above derivation of the pp $T$-matrix is elegant, this approximation was already well-known.
However, this formalism offers a new systematic path to include corrections on top of the $T$-matrix approximation through the irreducible vertex function $\tilde{\Gamma}$.
\titou{In the remainder of this section, vertex corrections to the self-energy and, in a second stage, to the irreducible pp propagator $\tilde{K}$ will be discussed.
Note that, in the context of traditional Hedin's equations, improving the self-energy without improving the effective interaction (and vice versa) has produced mixed results \cite{Shirley_1996,Schindlmayr_1998,Gruneis_2014,Maggio_2017b,Lewis_2019,Wang_2021a,Forster_2022a,Mejuto-Zaera_2022,Bruneval_2024}.}

The lowest-order term of the Gorkov irreducible vertex function arising at the first iteration is of first order in the effective interaction $T$ (see \SupInf).
Following Eq.~\eqref{eq:tilde_gamma_alt_hedin}, the component of interest in the normal phase ($U=0$) is given by
\begin{multline}
    \dfrac{T^\ee(3'7;7'1')}{2} \\
    \times \eval{[ G^\hh(74') G^\hh(47') - G^\hh(7'4') G^\hh(47) ]}_{U=0} = 0,
\end{multline}
which means that there is no self-energy correction of second order in $T$ in the number-conserving limit.
Hence, the first non-zero self-energy terms beyond Eq.~\eqref{eq:gt_selfnrj} are of third order in $T$.
This could have been anticipated as the $T$-matrix self-energy is exact up to second-order in the Coulomb interaction.

\begin{figure}
  \centering
  \includegraphics[width=0.3\linewidth]{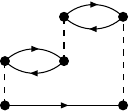}
  \caption{A third-order self-energy term arising through the second iteration of the pp Hedin equations. \label{fig:3rdorder_bubble}}
\end{figure}

To exemplify the possibility of going beyond Eq.~\eqref{eq:gt_selfnrj}, we report a third-order self-energy term that is non-zero in the absence of pairing fields.
This term is naturally obtained at the second iteration of the pp Hedin equations (see \SupInf) and reads
\begin{multline}
  \label{eq:3rdorder_selfnrj}
  - \ii^2 G(2'2^{++}) T(12;33') T(66';2'8) G(8^{--}8')  \\
  \times T(8'7;7'1') G(7'6') G(3'7) G(36).
\end{multline}
The lowest-order diagram in $v$ contained in Eq.~\eqref{eq:3rdorder_selfnrj} is represented in Fig.~\ref{fig:3rdorder_bubble}.
Equation \eqref{eq:3rdorder_selfnrj} is diagrammatically equivalent to the third-order $GW$ bubble diagram where the Coulomb interaction has been replaced by the effective interaction $T$.
This is fully analog to the screened ladder diagrams that arise through the vertex corrections to the $GW$ self-energy in conventional Hedin's equations \cite{Schindlmayr_1998,Mejuto-Zaera_2022}.

\begin{figure}
  \centering
  \includegraphics[width=0.3\linewidth]{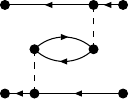}
  \caption{A second-order irreducible pp propagator term arising through the second iteration of the pp Hedin equations. \label{fig:2nd_order_pp_propag}}
\end{figure}

\titou{For the same reason as above, the lowest-order inner-vertex corrections are of second order in $T$.
Using the same irreducible vertex as the one giving rise to Eq.~\eqref{eq:3rdorder_selfnrj} (see \SupInf) leads to the following term for the irreducible pp propagator
\begin{multline}
  \label{eq:2nd_order_pp_propag_eq}
  -G(31')G(3'2')T(66';38)G(8^{--}8')G(7'6') \\
  \times \frac{T(8'7;7'3')}{2}[G(27)G(16)-G(26)G(17)].
\end{multline}
This term should be added to $K_0$ and used to compute a new effective interaction using Eq.~\eqref{eq:t_alt_hedin}.
Figure \ref{fig:2nd_order_pp_propag} displays the lowest-order diagram in $v$ contained in the above equation.
It corresponds to the propagation of two particles interacting through a second-order screened interaction.
The term derived in Eq.~\eqref{eq:2nd_order_pp_propag_eq} corresponds to a similar process where the bare Coulomb interaction has been replaced by the effective interaction $T$. 
This shows that the inner-vertex corrections introduce screening in $\tilde{K}$ and, hence, in the effective interaction.}

\titou{Including the self-energy term of Eq.~\eqref{eq:3rdorder_selfnrj} as well as the inner-vertex correction of Eq.~\eqref{eq:2nd_order_pp_propag_eq} would certainly be valuable in systems were pairing correlations are dominant but screening is non-negligible.
For example, pairing correlations are essential to qualitatively describe the \SI{6}{\eV} satellite of nickel but screening is also necessary for quantitative agreement \cite{Liebsch_1981,Springer_1998}.
In a different context, Pisani and co-workers have shown that going beyond the $T$-matrix approximation, in particular including screening, is essential to describe the physics of the BCS-BEC crossover in Fermi gases \cite{Pisani_2018a,Pisani_2018b}.
The ladder and bubble diagrams have also been considered simultaneously, through the FLEX approximation \cite{Bickers_1989a,Bickers_1989b}, to study superconductivity \cite{Takimoto_2004,Kitatani_2015}.}

\section{Conclusion}
\label{sec:conclusion}

In this work, we introduced a new system of equations for the one-body propagator $G$.
The lowest-order self-energy approximation coming from this set is the well-known pp $T$-matrix approximation, where $T$ is computed at the pp-RPA level.
Self-consistently iterating this set formally leads to a perturbative expansion of the self-energy with respect to the pp $T$-matrix.
This procedure parallels the self-energy expansion in terms of the screened interaction $W$ obtained through the conventional form of Hedin's equations.
Therefore, we refer to this new set as pp Hedin's equations.
More importantly, this framework allows us to derive, from first principles, vertex corrections to the $T$-matrix approximation.

The pp Hedin equations have been obtained by first deriving a closed set of equations for the Gorkov propagator in the presence of an external pairing potential and then 
taking the limit of a vanishing potential.
Indeed, the pp $T$-matrix interaction naturally appears when one seeks the response of an anomalous propagator to a pairing field.
Consequently, this derivation is thus more appropriately performed in the Nambu-Gorkov framework rather than considering solely the normal one-body propagator.

Starting from the pp Gorkov-Hedin equations, the simplest form of the irreducible vertex function leads to the generalized $T$-matrix self-energy introduced by Bozek to study superfluid nuclear matter \cite{Bozek_2002}.
This new functional derivative perspective brings complementary insight into Bozek's diagrammatic derivation.
For example, Bozek's $T$-matrix self-energy is not complete up to second-order in the Coulomb interaction and we show that these missing terms arise through the lowest-order vertex correction.

This lowest-order vertex correction to the self-energy, of second-order in $T$, turns out to be zero in the normal phase.
The first non-vanishing outer- and inner-vertex corrections are obtained by performing a second iteration of the pp Hedin equations and is thus of third order in $T$.
Diagrammatically, the self-energy term corresponds to the two-bubble $GW$ self-energy diagram where the bare Coulomb lines have been replaced by $T$.
Once again, a parallel can be drawn with conventional Hedin's equations, where the vertex function generates screened ladder self-energy diagrams.
\titou{The inner-vertex correction to the pp propagator has also been shown to include screening diagrams in the approximation.}

Because the first inner- and outer-vertex corrections in the normal phase are of second and third order in $T$, respectively, this approach is likely computationally too expensive in practice.
\titou{(See, for example, Ref.~\cite{Bruneval_2024} where it has been shown that computing the dynamical self-energy of second-order in $W$ leads to a drastic increase of the computational cost.)
In addition, considering only inner- or outer-vertex corrections has produced mixed outcomes in the $GW$ case \cite{Shirley_1996,Schindlmayr_1998,Gruneis_2014,Maggio_2017b,Lewis_2019,Wang_2021a,Forster_2022a,Mejuto-Zaera_2022,Bruneval_2024}. This might also be the case for the $T$-matrix approximation.}
Therefore, an alternative route might be to combine $W$ and $T$.
This has already been explored in various ways, for example, by replacing the Coulomb interaction with a screened interaction in ladder self-energy diagrams \cite{Liebsch_1981,Romaniello_2012,Zhukov_2005,Nabok_2021}.
The fluctuation exchange approximation of Bickers and coworkers involves summing the $GW$ and $T$-matrix channels (without double counting) \cite{Bickers_1989a,Bickers_1989b,Bickers_1991,Springer_1998}.
The Fadeev RPA \cite{Schuck_1973,Barbieri_2001,Barbieri_2007,Degroote_2011}, \titou{parquet theory \cite{Krien_2021}} and multi-channel Dyson formalisms \cite{Riva_2022,Riva_2023} constitute other alternatives to approximately couple the various scattering channels.
We believe that the Gorkov propagator might offer yet another way to combine them and is currently being investigated in our group.

Finally, note that this work focused on using anomalous quantities to compute the one-body propagator.
Therefore, a natural extension would be to consider these quantities within the two-body Bethe-Salpeter equation.
In particular, it can be shown that pairing propagators and anomalous self-energies offer a convenient framework to compute the kernel of the pp Bethe-Salpeter equation.
Work in this direction is currently underway and will be presented in a subsequent study.

\acknowledgements{The authors thank Abdallah Ammar and Carlos Mejuto-Zaera for insightful discussions.
  This project has received funding from the European Research Council (ERC) under the European Union's Horizon 2020 research and innovation programme (Grant agreement No.~863481).}

\bibliography{anom_propag}

\end{document}